\documentclass[conference,a4paper,twocolumn]{IEEEtran}
\IEEEoverridecommandlockouts
\usepackage{cite}
\usepackage{amsmath,amssymb,amsfonts}
\usepackage{algorithmic}
\usepackage{graphicx}
\usepackage{textcomp}
\usepackage{xcolor}
\usepackage{verbatim}
\usepackage{cleveref}
\usepackage{algorithm}
\usepackage{booktabs}
\usepackage{subfigure}
\usepackage{makecell}
\usepackage{multirow}
\usepackage{arydshln}
\def\BibTeX{{\rm B\kern-.05em{\sc i\kern-.025em b}\kern-.08em
    T\kern-.1667em\lower.7ex\hbox{E}\kern-.125emX}}
\IEEEsettopmargin{t}{2.2cm}
\IEEEsettopmargin{b}{4.63cm}

\begin{document}

\title{ Performance optimization and analysis of the unstructured discontinuous Galerkin solver on multi-core and many-core architectures \\
%{\footnotesize \textsuperscript{}}
\thanks{*Corresponding author:dengliang11@nudt.edu.cn.
	This work was supported by the National Numerical Wind Tunnel of China.}
}

\author{\IEEEauthorblockN{1\textsuperscript{st} Zhe Dai}
\IEEEauthorblockA{\textit{ Computational Aerodynamics Institute } \\
\textit{China Aerodynamics Research and Development Center }\\
Mianyang, China \\
daizhe$\_$cardc@163.com}
\and
\IEEEauthorblockN{2\textsuperscript{nd} Liang Deng*}
\IEEEauthorblockA{\textit{ Computational Aerodynamics Institute} \\
\textit{ China Aerodynamics Research and Development Center }\\
Mianyang, China \\
dengliang11@nudt.edu.cn}
\and
\IEEEauthorblockN{3\textsuperscript{rd} Yueqing Wang}
\IEEEauthorblockA{\textit{ Computational Aerodynamics Institute } \\
\textit{ China Aerodynamics Research and Development Center }\\
Mianyang, China \\
yqwang2013@163.com }
\and
\IEEEauthorblockN{4\textsuperscript{th} FangWang}
\IEEEauthorblockA{\textit{ State Key Laboratory of Aerodynamics } \\
\textit{ China Aerodynamics Research and Development Center }\\
Mianyang, China \\
wangfang@cardc.cn }
\and
\IEEEauthorblockN{5\textsuperscript{th} Ming Li}
\IEEEauthorblockA{\textit{ Computational Aerodynamics Institute } \\
\textit{ China Aerodynamics Research and Development Center }\\
Mianyang, China \\
liming@cardc.cn}
\and
\IEEEauthorblockN{6\textsuperscript{th} Jian Zhang}
\IEEEauthorblockA{\textit{  Computational Aerodynamics Institute } \\
\textit{China Aerodynamics Research and Development Center}\\
Mianyang, China \\
zhangjian@cardc.cn}
}

\maketitle

\begin{abstract}
The discontinuous Galerkin (DG) algorithm is a representative high order method in Computational Fluid Dynamics (CFD) area which possesses considerable mathematical advantages such as high resolution, low dissipation, and dispersion. However, DG is rather computationally intensive to demonstrate practical engineering problems.
This paper discusses the implementation of our in-house practical DG application in three different programming models, as well as some optimization techniques, including grid renumbering and mixed precision to maximize the performance improvements in a single node system. 
The experiment on CPU and GPU shows that our CUDA, OpenACC, and OpenMP-based code obtains a maximum speedup of 42.9x, 35.3x, and 8.1x compared with serial execution by the original application, respectively. 
Besides, we systematically compare the programming models in two aspects: performance and productivity. Our empirical conclusions facilitate the programmers to select the right platform with a suitable programming model according to their target applications.
\end{abstract}

%However the measurement on performance and productivity is still a problem especially for practical engineering code which is sensitive to the cost of maintenance and development.

\section{Introduction}

The high-order scheme receives considerable attention in Computational Fluid Dynamics (CFD) due to its low numerical dissipation for Large-Eddy Simulation (LES)\cite{grube2007assessment} and the capability to capture the fine-grained structure in complicated flows.
Discontinuous Galerkin (DG)\cite{reed1973triangular} is a popular algorithm that possesses good mathematical properties in conservation, stability, and convergence in complex flow problems.
However, DG suffers from severe computing overhead comparing with the low-order scheme. 
For example, solving the Euler equation with three-dimensional fourth-order DG method needs one hundred variables per grid element\cite{LI2015628,ZHANG20121104}, while the only single-digit variable is required in low-order methods, not to mention the additional volume and area integral computations in every grid element, which is not needed at all in low-order method. 

To make computationally expensive DG algorithm available in practical engineering, many researchers have made considerable achievements on both GPU and CPU haredwares.
Xia accomplishes the unstructured grid DG algorithm based on OpenACC Programming model\cite{xia2014openacc}.
Pazner employed a parallel-in-time strategy to compute the Runge-Kutta stages simultaneously under DG discretizations\cite{PAZNER2017700}.
Duan applied total variation diminishing Runge–Kutta scheme coupled with the multigrid strategy to improve the parallel efficiency of DG method on CPUs\cite{Purpose2021}.
Maurice presented a parallel computing strategy for a hybridizable discontinuous Galerkin nested geometric multigrid solver on many-core hardware, attaining 80$\%$ of peak performance for higher order polynomials\cite{fabien2019manycore}.
Kronbichler integrated optimized DG code into a scalable framework of five supercomputers using both CPU and GPU hardwares\cite{KronbichlerSC21}.

\begin{comment}
Goedel solves the electromagnetic radio frequency problem in four GPUs, achieving a speedup of 36.7 compared with four AMD CPUs\cite{goedel2009gpu}.
Chan solves the time domain acoustic equation for mixed grids on GPUs\cite{chan2016gpu}, evaluating the computing cost of different grids in Flops, and memory bandwidth metrics.
Klockner develops the nodal electro-magnetic field algorithm on CUDA\cite{klockner2009nodal}, achieving a speedup of 60x. Mu simulates the seismic wave propagation based on MPI+CUDA, obtaining a speedup of 14.9\cite{mu2013accelerating}.
\end{comment}

The former studies usually focus on the algorithm optimization and performance achievement with either CPU or GPU. 
However, few articles consider the performance portability across multi-core CPU and many-core accelerators, as they are generally used in high performance computing (HPC) system.
The latest TOP500 list shows that six out of the top ten supercomputers utilize many-core accelerators, this trend brings not only a good opportunity to accelerate DG algorithm on GPU, but also barriers in programming to migrate code on diverse computing architectures.

Therefore, this work aims to assess three programming models, namely OpenMP, OpenACC, and CUDA on both multi-core CPU and many-core accelerator, in the context of an unstructured grid high-order DG application called HOUR2D. 
The aspects to be compared are the implementational effort and performance evaluation, as well as the optimizations during code migration.
Finally, we evaluate the perforamnce and portability to give insights on appropriate language selection for different purposes.
The work provides novelty from several perspectives:

\addtolength{\topmargin}{0.4cm}

%Performance portability is an increasingly important topic for scientific applications as the hardware tends to be complexity and heterogeneous and the programming model is diversifying too, there's a great demand for the applications on refactoring the code to achieve higher performance in the updated high performance computing system, however porting the code from practical engineering applications is usually a laborious task due to the bulky amount of code.
%Particularly, we focus on explicit Runge-Kutta Navier–Stokes and Euler solvers, as turbulent flow physics drives the algorithmic choices towards these schemes which avoid severe CFL number restrictions associated with explicit schemes.

\begin{itemize}
	\item	We give a high-performance migration of HOUR2D application in OpenMP, OpenACC, and CUDA, respectively, the parallelization in different programming models is elaborately chosen to adapt unique data structures and computing formats.
	\item   We perform the grid renumbering method to optimize irregular memory access issue caused by the unstructured grid, then carry out the mixed floating-point method to improve computational performance, at last, the Unified Memory(UM) method are tested in OpenACC for workload reduction.
	\item	We evaluate the performance and productivity, firstly the speedup of three programming models is listed using six different grid meshes, then some insights are indicated by using the roofline model, at last, a comprehensive assessment is displayed together with productivity metric.
	%according to the performance indicator evaluted by profiler tools and quantification metric.
\end{itemize}

This work is organized as follows:
Section 2 gives a brief review of the mathematical method and typical data pattern used in HOUR2D.
Section 3 displays the implementation to achieve high performance for three prrgramming languages and the evaluation approaches.
Section 4 presents the experiment of performance and productivity in HOUR2D code.
Section 5 summarizes our whole research and gives insights.
\section{HOUR2D Mathematical Method}

HOUR2D can solve steady or unsteady field Navier-Stoke and Euler problems by using the explicit Runge-Kutta high-order DG algorithm, and it's mainly composed of three parts: preprocessing, flow field calculation, and output postprocessing.
%as shown in \cref{fig1}.
\begin{comment}
\begin{figure}[htbp]
	\centerline{\includegraphics{fig/fig1.png}}
	\caption{The main compositions of HOUR2D }\label{fig1}
\end{figure}
The preprocessing includes calculation parameters and grid processing, the calculation part includes flow field initialization and time advance step, and the postprocessing is the calculation result processing and output, where the time iteration step is the absolute performance hotspot of the software.
\end{comment}
The application uses the same high-order DG numerical method to discretize the Euler equation and Navier-Stokes equation, the governing equation, and the detailed numerical discretization process are as follows.

\subsection{ Governing Equation }
In the Cartesian coordinate system, the two-dimensional conserved form Navier-Stokes or Euler equations without regard to volume force and external heating are:
\begin{equation}
\label{eq1}
\frac{\partial Q }{\partial t}+\frac{\partial E}{\partial x}+\frac{\partial F}{\partial y}=ivis\cdot (\frac{\partial _{E_v}}{\partial x} + \frac{\partial _{F_v}}{\partial x})
\end{equation}
where $Q$ is conserved variable, while ($E,F$)/($E_v,F_v$) are inviscid/viscous flux in $x$,$y$ direction respectively, \cref{eq1} is Euler equation  when $ivis=0$, otherwise is NS equation when $ivis=1$. The detailed formation is
\begin{equation}
\begin{split}
Q=\begin{bmatrix}
\rho  \\
\rho \mu  \\
\rho \nu \\
e\\
\end{bmatrix},E=\begin{bmatrix}
\rho \mu \\
\rho \mu^{2} + p\\
\rho \mu\nu \\
(e+p)\mu\\
\end{bmatrix},F=\begin{bmatrix}
\rho\nu\\
\rho \nu\mu\\
\rho\nu^2+p\\
(e+p)\nu\\
\end{bmatrix}  \\
\end{split}
\end{equation}

\begin{equation}
\begin{split}
E_{\nu }=\begin{bmatrix}
0 \\
\tau _{xx}\\
\tau _{xy}\\
\mu \tau _{xx}+\nu \tau _{xy}-q_{x}\\
\end{bmatrix},F_{\nu }=\begin{bmatrix}
0 \\
\tau _{yx}\\
\tau _{yy}\\
\mu \tau _{yx}+\nu \tau _{yy}-q_{y}\\
\end{bmatrix}
\end{split}
\end{equation}
where $\rho$ is the density, and $\mu$,$\nu$ is the velocity in $x,y$ direction,$p$ is the pressure, and $e$ is total fluid energy per unit volume written as:
\begin{equation}
 e=\rho \left [ U+\frac{1}{2}(\mu ^{2} + \nu ^{2}) \right ]=\frac{p}{\gamma -1}+\frac{1}{2}\rho (\mu ^{2} + \nu ^{2})
\end{equation}
while $U$ is the internal energy per unit mass of fluid:
\begin{equation}
U=C_{\nu }T=RT/\left ( \gamma -1 \right )=p/\rho /\left ( \gamma -1 \right )
\end{equation}
where $C_{\nu }=R/(\gamma-1)$ is specific heat capacity at constant volume,$T$ is the temperature obtained from state equation $p=\rho R T$,$R$ is gas constant,$\gamma$ is specific heat ratio.$\tau_{xx},\tau_{xy},\tau_{yx},\tau_{yy}$ are the weights of viscous stress tensor, Stokes' hypothesis for a Newtonian, isotropic fluid can be written as:
\begin{equation}
\tau=\begin{bmatrix}
\tau _{xx} & \tau _{xy} \\
\tau _{yx} & \tau _{yy} \\
\end{bmatrix} =\mu \begin{bmatrix}
2/3(2\mu _{x} - \nu _{y}) &  \mu _{y}+\mu _{x}\\
\mu _{y}+\mu _{x} & 2/3(2\nu _{y} - \nu _{x}) \\
\end{bmatrix}
\end{equation}
where $\nu$ is dynamic viscosity coefficient.
\subsection{ DG Discretization}
The governing equation in \cref{eq1} can be abbreviated as
\begin{equation}
Q_{t}+ \bigtriangledown \cdot \overrightarrow{F}\left ( Q \right )- \bigtriangledown \cdot \overrightarrow{F_{\nu }}\left ( Q,\bigtriangledown Q \right )=0
\end{equation}
compared with the second-order finite volume (FVM) discretization method, all the physical quantities of every element obey high-order polynomial distribution in DG discretization way, which means the corresponding distribution polynomial coefficients are needed, scilicet, degrees of freedom.
\begin{equation}
\label{eq10}
Q\left ( t,x,y \right ) = \sum_{j}c_{j}\left ( t \right )b_{j}\left ( x,y \right )
\end{equation}
 $c_j$ in \cref{eq10} is the physical quantity coefficient in element which varies over time, $b_j$ is basis function, then introducing auxiliary variables based on \cref{eq10}:

\begin{center}
\begin{equation}
\begin{split}
  \overrightarrow{Z}\left ( Q \right ) =\bigtriangledown Q \\
  Q_{t}+\bigtriangledown \cdot \overrightarrow{F}\left ( Q \right )-\bigtriangledown \cdot \overrightarrow{F_{v}} ( Q,\overrightarrow{Z} )=0
\end{split}
\end{equation}
\end{center}
afterward multiplying again by basis function $b_j$, at last, we get a discrete form of the discontinuous finite element using integration by parts :
\begin{equation}\label{eq11}
\begin{split}
\int _{\Omega }\overrightarrow{Z}b_{j}d\Omega = \\
-\int _{\Omega }Q\cdot \bigtriangledown b_{j}d\Omega + \int _{i}\hat{Q}\left ( Q^{-} , Q^{+}\right )\overrightarrow{n}b_{j}d\Gamma \\
+\int _{b}\hat{Q^b}\left ( Q^{-} , Q^{b}\right )\overrightarrow{n}b_{j}d\Gamma \int _{\Omega }Q_{t}b_{j}d\Omega \\
- \int _{\Omega }\left(\overrightarrow{F}\left ( Q \right )- \overrightarrow{F_{v}}\left ( Q,\overrightarrow{Z} \right )\right)\cdot \bigtriangledown b_{j}d\Omega \\
+\int _{i}\overrightarrow{H_{c}}\left ( Q^{-},Q^{+},\overrightarrow{n}\right )b_{j}d\Gamma \\
+\int _{b}\overrightarrow{H^b_{c}}\left ( Q^{-},Q^{b},\overrightarrow{n}\right )b_{j}d\Gamma\\
-\int _{i}\overrightarrow{H_{v}}\left ( Q^{-},\overrightarrow{Z}^-,Q^+,\overrightarrow{Z}^+,\overrightarrow{n} \right )b_{j}d\Gamma \\
-\int _{b}\overrightarrow{H^b_{v}}\left ( Q^{-},\overrightarrow{Z}^-,Q^b,\overrightarrow{Z}^b,\overrightarrow{n} \right )b_{j}d\Gamma \\
=0
\end{split}
\end{equation}

%\addtolength{\topmargin}{0.4cm}

The spatial integral in \cref{eq11} is a Gaussian Numerical Integral of which the subscripts $\Omega, i,b$ represents the element integral, the interior surface integral, and the boundary surface integral respectively, 
while the superscript$-,+,b$ represents the physical quantity inside the element, outside the element, and at the boundary conditions.
$H_c$ denotes inviscid flux which is in Roe format\cite{roe1981approximate}, besides $H_\nu$ is viscous flux in BR1 format\cite{cockburn1998runge}.

\section{ Performance Implementation}

\begin{figure}[htbp]
	\centerline{\includegraphics[width = 0.4\textwidth]{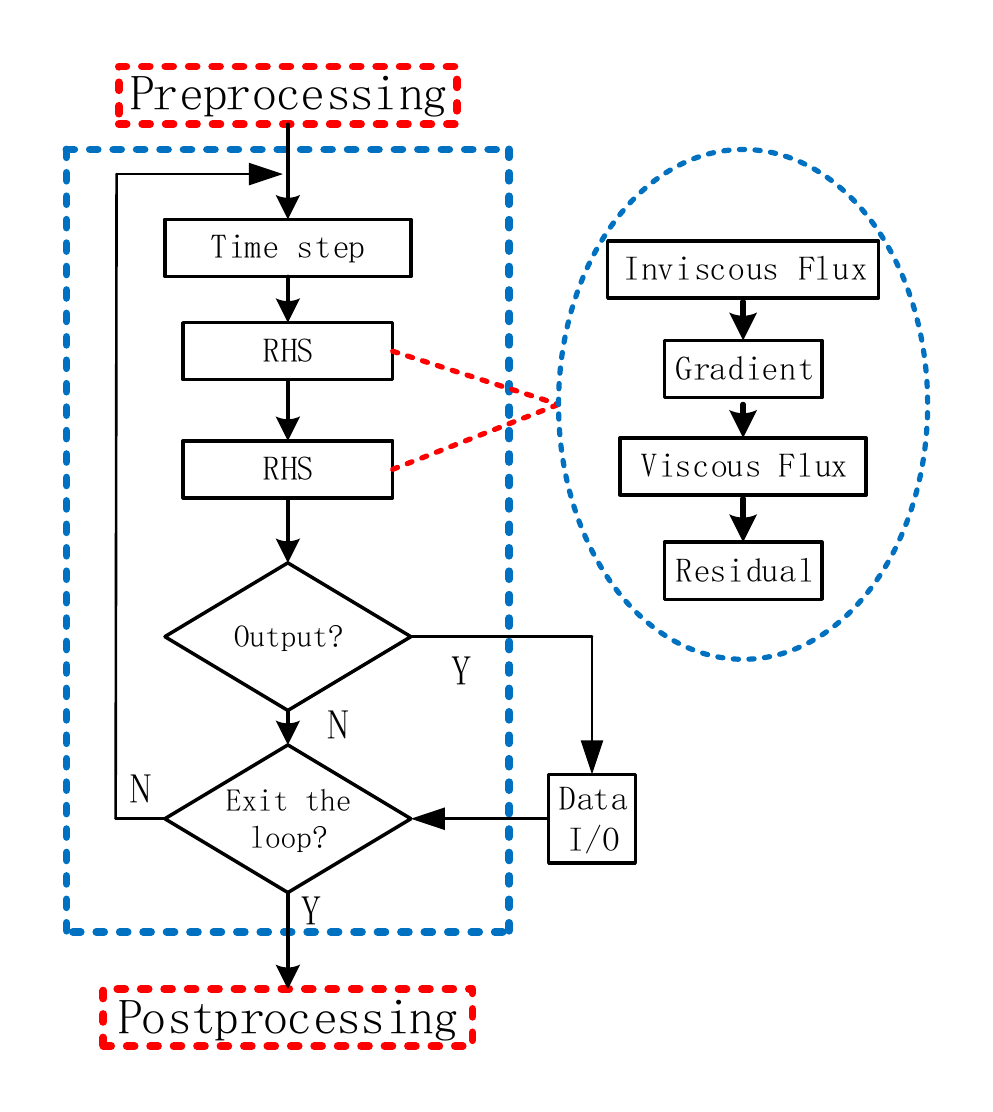}}
	\caption{The major work flows for HOUR2D application. }\label{fig2}
\end{figure}

Time iterations is the hotspot which occupies more than 95\% of total running time which is shown detailedly in \cref{fig2}, including time step, right-hand-side(RHS), and file output, the RHS procedure involves inviscid flux, gradient, viscous flux, and residual calculation.
The RHS step is counted twice in every time step because of the two-order explicit Runge-Kutta method, and it consumes approximately 90\% of total running time, so it is the computationally intensive part of the whole application indeed, therefore the primary concern for performance optimization is RHS. 
\subsection{ Data Structure and Computing Formats}

\begin{figure}[htbp]
	\centerline{\includegraphics[width = 0.4\textwidth]{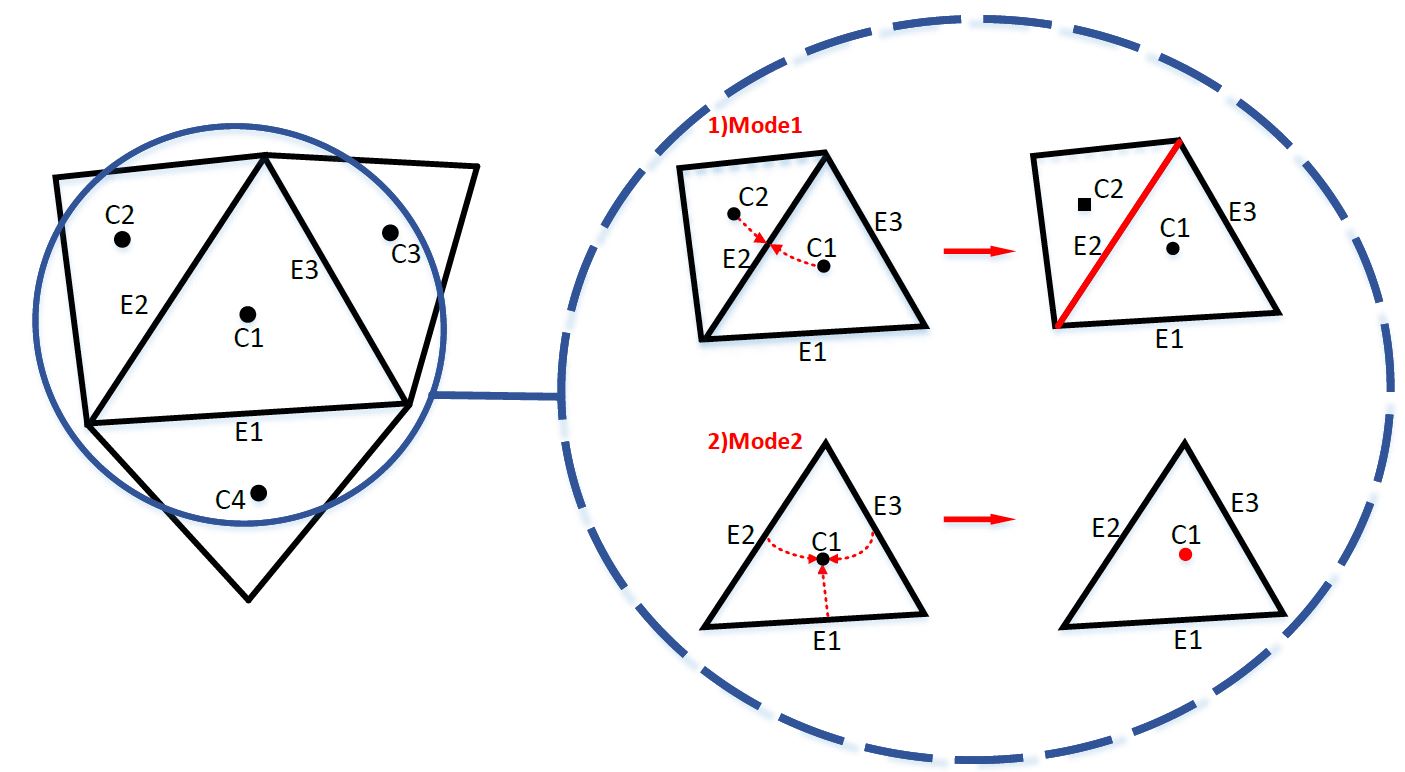}}
	\caption{Data structure and major computational modes. $C$ is the body center of the cell, and $E$ is the face unit where the data is stored in AOS format. The dotted box shows two major modes in our computations, the red dotted arrow means the data access flows in memory.}\label{fig3}
\end{figure}

The geometric topology of HOUR2D's unstructured mesh consists of vertices, elements, and cells as shown in \cref{fig3}, $E_i$ is the face unit in which stores the flux variable, $C_i$ is the cell unit in which stores the intermediate variables that include density, velocity, and pressure, besides the above data is organized as the array of structure(AOS) format.

We display two major computational modes that affect memory access pattern and parallelization format as the dotted box shows.
The first one indicates the flux computation which is parallelized through elements, for every time step, the program reads the intermediate variables from neighbor cell units, then writes back to the current element unit after calculations. 
The other one is the physical quantity calculations that are stored in cells, which requires flux variable around adjacent cells at the present step, data race occurs if the element parallelization model is still utilized.

\begin{comment}
We avoid the potential conflicts by applying the cell-based loop instead of the element-based loop as ~\cref{alg1} shows, 
the $ncell$ is the number of total cells, $NumOfElem$  shows the neighbor element number at the current cell, and $geo$ is the array of the geometric variable. 
In this way, the associated loops can be parallelized directly with the cost of redundant calculations in every nearby element of data.

\begin{algorithm}[htp]
	\caption{ The Algorithm of Cell-based Loop }
	\label{alg1}
	\begin{algorithmic}[1]
		%seqence
		\FOR { nc=0 : ncell }
		\STATE {    face $\leftarrow$ NumOfElem[nc]; }
		\FOR { nf=0 : face  }
		\STATE { f $\leftarrow$ cell[nc].ofElem[nf];       }
		\STATE { flux $\leftarrow$ element[f].flux;        }
		\STATE { cell[nc].rhs += geo[nc]$*$flux;  }
		\ENDFOR
		\ENDFOR
	\end{algorithmic}
\end{algorithm}

\end{comment}
\subsection{Grid Renumbering }

The second computing formats mentioned in the former subsection involves the data access aroud the neighbor cells in every central cell.
However, the number of these adjacent cells are not naturally continuous as the toplogy of unstructured grid is irregular.
Hence, resulting in indirect memory access patterns when iterating over the cells of the unstructured grid for the flux, gradient, and residual computations.
%The operations can be summarized as gather and scatter format which manipulates across large random strides in memory depending on the mesh topology and algorithms used in the grid generation stage. 

%\addtolength{\topmargin}{0.4cm}

\begin{figure}[htbp]
	\centerline{\includegraphics[width = 0.4\textwidth]{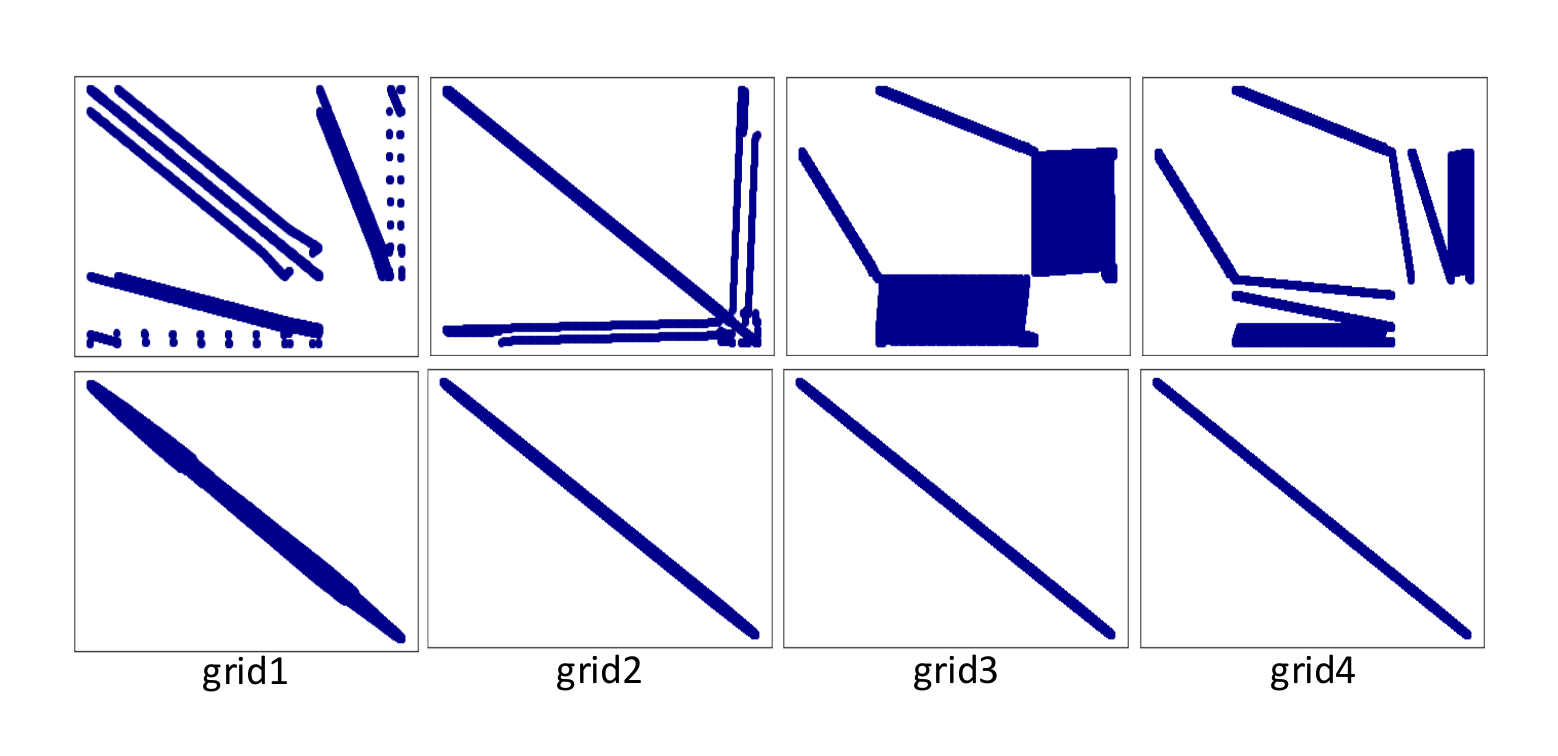}}
	\caption{ The intuitive illustration for bandwidth reductions of four computional grids. The upper figures show the original adjacency, while the lower figures are adjacency after applying RCMK. }\label{fig4}
\end{figure}

To relieve the irregular memory access issue, we reorder the grid to reduce the cache miss and improve the data locality with Reverse Cuthill Mckee (RCMK) algorithm \cite{cuthill1969reducing}.
The RCMK is developed to minimize the bandwidth in the adjacent graph by assembling the non-zero values as close as possible to the main diagonal\cite{burgess1997renumbering}.
So we reorder the topological adjacency derived from the computational grid to align the cell numbers.
The optimization result is shwon in \cref{tab1}, it displays the bandwidth of four unstructured grids with different number of cells.
The last two columns are the bandwidth of cell-based adjacency matrix which shows the degree of aggregation around the diagonal, the bandwidth is greatly reduced according to the results of all four grids, it can be infered that bandwidth reduction obtains 2-3 orders of magnitude.
A more detailed plot about the cell aggregation in shown in \cref{fig4}, it exhibits the fairly good result by comparing the adjacency before and after applying RCMK.

\begin{table}[htbp]
	\centering
	\caption{ The bandwidth of different computational grids when using RCMK algorithm. The element is the number of grid cells, bandwidth shows the max aggregation of adjacent elements, The bandwidth is greatly reduced after applying RCMK. }
	\vspace{0.1cm}
	\small
	\begin{tabular}{ p{1cm}p{1.8cm}p{1.8cm}p{1.8cm} }
		\toprule
		\makecell[c]{Grids}          &     \makecell[c]{ Number \\ of cell}             &     \makecell[c]{Original \\ bandwidth}    &   \makecell[c]{RCMK \\ bandwidth}   \\
		\hline
		\makecell[c]{Grid1} &   \makecell[c]{2688}     &      \makecell[c]{1427}        &      \makecell[c]{69}         \\
		\makecell[c]{Grid2} &   \makecell[c]{4032}     &      \makecell[c]{3992}        &      \makecell[c]{35}         \\
		\makecell[c]{Grid3} &   \makecell[c]{16128}    &      \makecell[c]{12410}       &      \makecell[c]{67}         \\
		\makecell[c]{Grid4} &   \makecell[c]{64512}    &      \makecell[c]{49016}       &      \makecell[c]{131}        \\

		\bottomrule
	\end{tabular}
	\label{tab1}
\end{table}

Following the renumbering, the cell is arranged in the order of monotonical index, and a consecutive rank is performed on the element index according to the first constant cell index, so the second index reference will be visited in ascending order.
The staple benefits of the optimizations above can be summarized as enhancing both spatial and temporal locality particularly when cells are referenced contiguously in memory and exploiting the available memory bandwidth. 
What's more, the renumbering technique is executed only once before the time iteration starts for every process, hence it has a negligible effect on the whole application execution not to mention the time iteration is usually a fairly big number.
The overall performance improvement by using the grid renumbering method is about 20\%-30\%.

\subsection{ Mixed Precision }

The physical variables use the double-precision (DP) format to maintain good convergence and computational precision through CFD simulations compared with single-precision (SP), however, it increases the amount of memory data required for every floating-point operation, and the memory wall issue is further aggravated and restricts the CFD performance efficiency, under the conditions that off-chip memory bandwidth is considered to be the constraining resource in system performance for the foreseeable future\cite{patterson2004latency}.
The mathematical evidence proves that there's no need to use DP format in the all physical parameters or the whole calculation procedures, even if taking into account the constraint of given computational precision\cite{abdelfattah2021survey,tisseur2001newton},
so CFD application turn to use lower floating-point precision variables to achieve higher performance, naturally some mixed-precision(MP) strategies emerged\cite{kampolis2010cfd,gomes2021sustainable}.

\begin{figure}[htbp]
	\centerline{\includegraphics[width = 0.4\textwidth]{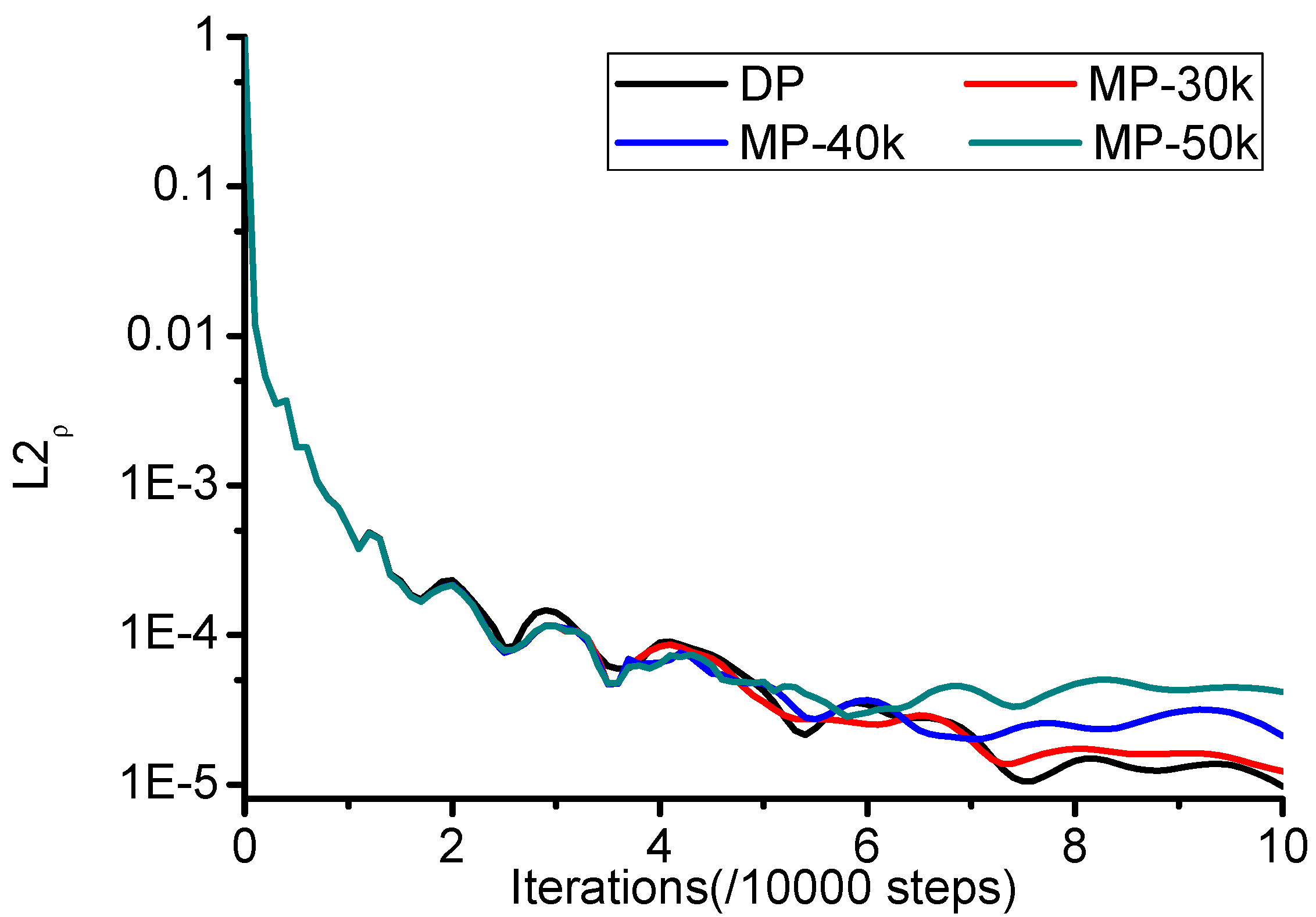}}
	\caption{ Mixed precision approach verification using NACA grid. The comparison of the residual convergence in MP methodology with different time steps allocated in SP and DP, for example, MP-30k means the initial 30000 iterations are executed in SP format, the rest of the iterations are computed in DP. }\label{fig5}
\end{figure}

One of the methods is called the iterative refinement approach, at first, it employs SP parameters in previous iterations to improve the computational performance, then uses DP format immediately when the convergence rebound occurs, so convergence, precision, and high performance are guaranteed at the same time. 
We verify this approach in HOUR2D by using a NACA airfoil mesh grid as the \cref{fig5} shows, 
%\cref{fig5}a is the pneumatic pressure CP, the red line is MP format with SP in 50 thousand steps, and DP in 50 thousand steps, the blue line is the benchmark computed in DP.
\cref{fig5} illustrates the $L_2$ residual convergence, in which MP-30k stands for the initial 30 thousand steps are computed in SP format, while the rest of the steps are computed in DP, it's obvious that the precision of the output result decline as the SP step increases.

\begin{comment}
\begin{figure}[htbp]
	\centering
	\subfigure[]{\includegraphics[width=0.40\textwidth]{fig/fig5-1.pdf}}
	\hspace{2cm}
	\subfigure[]{\includegraphics[width=0.40\textwidth]{fig/fig5-2.pdf}}
	\caption{ Mixed precision approach verification using NACA grid. (a) The comparison of pneumatic pressure metric between DP and MP with the total 100 thousand time steps. (b) The comparison of the residual convergence in MP methodology with different time steps allocated in SP and DP, for example, MP-30k means the initial 30000 iterations are executed in SP format, the rest of the iterations are computed in DP.  }\label{fig5}
\end{figure}
\end{comment}

%\IEEEsettopmargin{b}{5.0cm}

We assess the performance improvement of our MP method after the correctness verification with two grid examples as \cref{tab1} shows. 
The experiment result is shown in \cref{tab2}, in which the total time steps are 100 thousand, too.
DP shows the benchmark of our test, the run time decreases as the SP step increases.
The last row SP indicates the most optimal improvement with no DP format variables are used at all, we can conclude from the table that the MP method achieves 15\% performance improvement at most compared with the DP format.
However, the SP format is only comparable in performance, as the computational precision is not strictly guaranteed to meet the constraint in diverse grids from mathematical point of view, the SP format are usually not used in engineering computations, so the most optimal speedup is usually only a theoretical conclusion.

\begin{table}[htbp]
	\centering
	\caption{ Clock time of our mixed precision approach running on two grids. The DP means the double precision calculation, and SP is single precision format, the MP-20k means the starting 20000 iterations are executed in SP format, followed by the rest of the iterations computed in DP.}
	\vspace{0.1cm}
	\small
	\begin{tabular}{ p{1.5cm}p{1.5cm}p{1.5cm} }
		\toprule
		\makecell[c]{Time}  &   \makecell[c]{Grid1}  &  \makecell[c]{Grid2}    \\
		\hline
		\makecell[c]{DP}    &   \makecell[c]{51.87}    &       \makecell[c]{151.91}   \\
		\makecell[c]{MP-20k} &   \makecell[c]{49.68}    &       \makecell[c]{145.83}   \\
		\makecell[c]{MP-50k} &   \makecell[c]{48.30}    &       \makecell[c]{140.90}   \\
		\makecell[c]{MP-80k} &   \makecell[c]{46.68}    &       \makecell[c]{135.95}   \\
		\makecell[c]{SP}    &   \makecell[c]{43.87}    &       \makecell[c]{131.40}   \\
		\bottomrule
	\end{tabular}
	\label{tab2}
\end{table}
%It's worth noting that the rebounce differs from the grid, and usually acknowledged in practical. 

\subsection{ Heterogeneous Implementations  }

We port HOUR2D code using both Compute Unified Device Architecture(CUDA) and OpenACC on Nvidia GPU to get an apparent comparison in performance and productivity. CUDA usually gains a better performance than OpenACC due to the fine-grained manipulation of hardware resources within kernels, however, this generally requires the time-consuming task of code revising to make use of the resources for developers. 
On the contrary, OpenACC is a declarative model using compiler pragmas which is not as effective as CUDA in computing performance.

The minimization of data transmission between CPU and GPU hardware is at prior consideration in HOUR2D as it's an expensive operation over the time iterations, so the covered data is transported to GPU memory in the grid initialization stage to avoid the potential bottleneck caused by the frequent data transfer.
We accommodate the grid data with the appropriate component from the GPU memory hierarchy like shared memory, constant memory, and global memory to achieve better memory access performance.
For example, HOUR2D high-order algorithm needs 393 geometrical constants to accomplish element-integral and cell-integral in each time iteration step, so we place it in the constant memory to reduce memory transactions launched by thread warp.

As for the code migration, we port the whole iteration code into GPU as the blue dotted box shows in \cref{fig2}, on the one hand, the parallelization in $RHS$ step can be implemented naturally in the way of \cref{fig3} shows without any data racing, the element-based or cell-based loop can be mapped to the one-dimensional thread in GPU directly.

On the other hand, the time step computation involves the minimal value of all local time steps in cell data, making data racing if it's executed in direct parallelization.
To achieve effective performance in entire computing procedures, we use GPU reduction to get minimal value by taking advantage of shared memory which enables synchronization within thread blocks. Furthermore, we design the consecutive reduction method to realize robust reduction under the restriction of max thread number in GPU block.

\begin{comment}
the detailed algorithm is shown in \cref{alg3}, the input parameters include time step array $StepArray$, total thread number $nThread$, threads per block $thPerBlk$ and the reduction times $k$, the reduction kernel function $Reduce()$.
\begin{algorithm}[htp]
	\caption{ The pseudo-code of continous reduction to get minimal time step }
	\label{alg3}
	\begin{algorithmic}[1]
		\REQUIRE
		StepArray,nElem,thPerBlk,k
		\ENSURE
		minTime
		\FOR { it = 0 : k }
		\STATE {    nThread $\leftarrow$ CalThread(it,thPerBlk); }
		\STATE {    grid.x $\leftarrow$ (nThread + block.x-1)/block.x;  }
		\STATE {    Reduce $\ll$ grid,block $\gg$ (StepArray, nThread);   }
		\ENDFOR
		\STATE {    $minTime$ $\leftarrow$ StepArray[0]; }
	\end{algorithmic}
\end{algorithm}
\end{comment}

Porting the code by using OpenACC is much more convenient than CUDA by using compiler pragmas to parallelize compute-intensive code, for example, the time step reduction mentioned above only needs to use $\#pragma \ acc \ parallel \ reduction$ instead, and the specified process will be done by PGI compiler, but the performance also highly depends on the data layout in PGI compiler and the effectiveness of code generation.

UM is a pragmatic technology that can be used by CUDA 6.X or higher version, it manages the data between CPU and GPU independently which was relied on programmers, hence explicit function calls and data migration can be decreased evidently.
We adopt the UM method during heterogeneous code migration by simply adding the $managed$ compilation option so that memory space between host and device is established.

UM leads to possible performance loss due to the additional memory transmission and memory page fault\cite{li2015evaluation}, which is also observed in our OpenACC code, generally, the performance loss decreases as the grid size increases, eventually stabilizing below 3\% especially when the grid's element exceeds sixteen thousand, this may be caused the intensification of discontinuous memory access as grid size increases.

\section{Experiments and Analysis}

\begin{figure*}[htbp]
	\centering
	\subfigure[]{\includegraphics[width=0.33\textwidth]{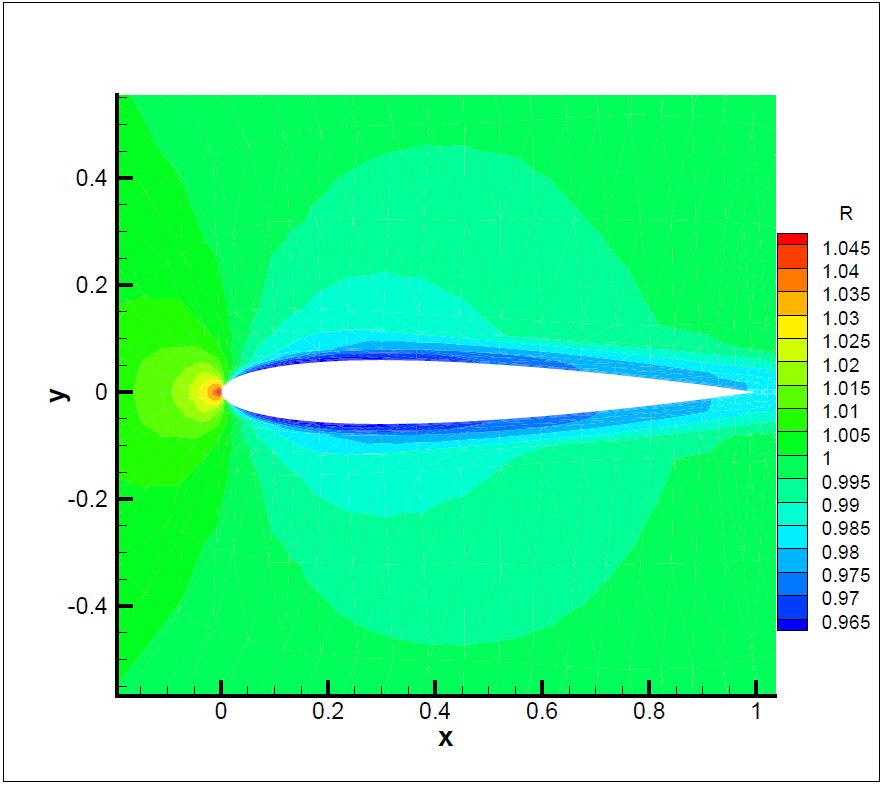}}
	\hspace{0.6cm}
	\subfigure[]{\includegraphics[width=0.33\textwidth]{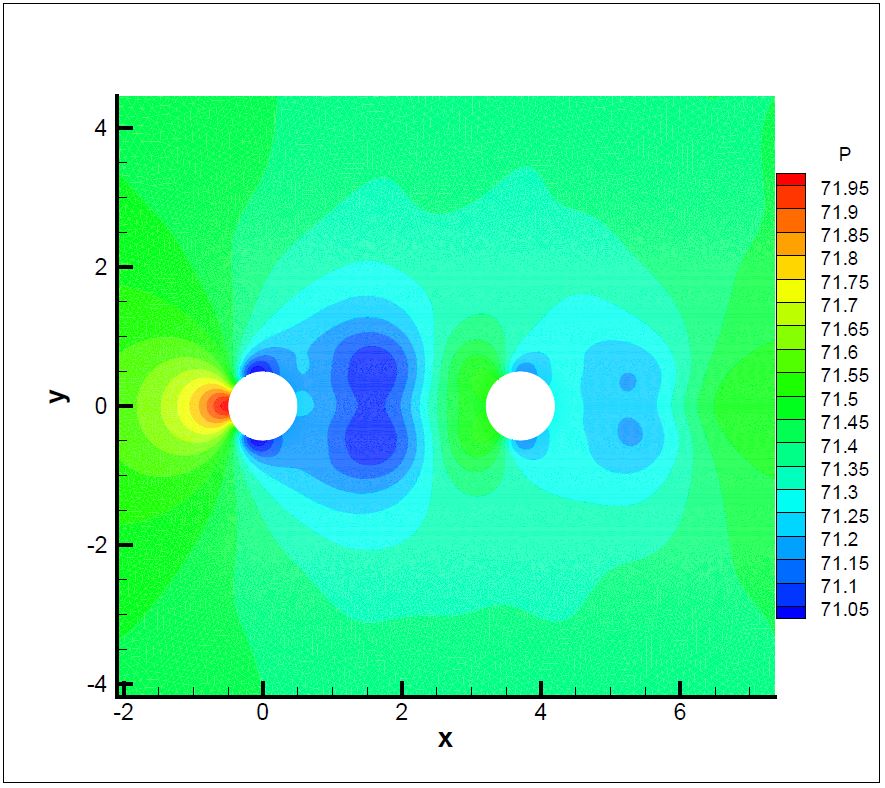}}
	\caption{ Streamline diagram of the real computing results. (a) The density of grid2, with Mach number=0.3, Reynolds number=5000 at 400 thousand time steps . (b) The pressure of double cylindrical, with Mach number=0.1, Reynolds number=100 at 400 thousand time steps.}\label{fig7}
\end{figure*}
\begin{comment}
\begin{figure}
	\centerline{\includegraphics[width=0.30\textwidth]{fig/fig7-1.jpg}}
	\caption{ The run time of OpenMP-based code with six different grid meshes }\label{fig7}
\end{figure}
\end{comment}
The HOUR2D code is fulfilled at the same level of optimization in three different programming models: OpenMP, OpenACC, and CUDA, while the computing hardware includes a computing node consisting of an Intel Xeon E5-2698 @2.20GHz CPU and an Nvidia A100 GPU, as for compiler toolchain, we use $GCC-5.4.0$ for CPU code with optimization flag $-O3$, and the GPU compiler $NVCC-10.2.89$ for CUDA and OpenACC.
The computational grid is illustrated in \cref{tab3}  which includes two steady NACA airfoil grids(grid1 and grid2), four unsteady front step grids (grid3-grid6), and an unsteady flow around double cylindrical(grid7). The latter grid is four times larger than the former one for those possess same aerodynamic shape, the last column is the grid type of every element.

\begin{table}[htbp]
	\centering
	\caption{ Detailed information of computational grids used in subsequent experiments. Nodes and cells show the computational scale, the difference of grid shape reflects the number of faces in cell. }
	\vspace{0.1cm}\small
	\small
	\begin{tabular}{ p{1cm}p{1.5cm}p{1.5cm}p{2cm} }
		\toprule
		\makecell[c]{Grid}   &   \makecell[c]{Nodes}    &   \makecell[c]{Cells} &   \makecell[c]{Grid shape}         \\
		\hline
		\makecell[c]{grid1}  &   \makecell[c]{378}      &   \makecell[c]{672}      &   \makecell[c]{triangular}    \\
		\makecell[c]{grid2}  &   \makecell[c]{1428}     &   \makecell[c]{2688}     &   \makecell[c]{triangular}    \\
		\makecell[c]{grid3}  &   \makecell[c]{4193}     &   \makecell[c]{4032}     &   \makecell[c]{quadrilateral} \\
		\makecell[c]{grid4}  &   \makecell[c]{16449}    &   \makecell[c]{16128}    &   \makecell[c]{quadrilateral} \\
		\makecell[c]{grid5}  &   \makecell[c]{65153}    &   \makecell[c]{64512}    &   \makecell[c]{quadrilateral} \\
		\makecell[c]{grid6}  &   \makecell[c]{259329}   &   \makecell[c]{258048}   &   \makecell[c]{quadrilateral} \\
		\makecell[c]{grid7}  &   \makecell[c]{248215}   &   \makecell[c]{495432}   &   \makecell[c]{quadrilateral} \\
		\bottomrule
	\end{tabular}
	\label{tab3}
\end{table}

\subsection{ Evaluation Approach }
As the main hotspot of HOUR2D is time iteration that performs almost the same operation, and tens of thousands of time steps are usually required to reach a convergent result in practical CFD simulation, the preprocessing and postprocessing parts should be removed for the seek precise evaluation. 
We apply Dual-phase measurement\cite{che2014microarchitectural} method to exclude the non-primary impacts during the performance experiments.
Specifically, the application is tested twice with different iterations, the subtraction data between two iterations shows the difference of exact iteration computations.
Hence, the performance metric like Flops, DRAM bytes and run time of the can be computed precisely in this way.

\begin{comment}
\begin{figure}[htbp]
	\centerline{\includegraphics[width=0.40\textwidth]{fig/fig6.pdf}}
	\caption{ Illustration of dual-phase measurement method. }\label{fig6}
\end{figure}
on the performance evaluation basis of HOUR2D code porting from the same desktop version to three different programming models, we try to give a performance portability quantification of HOUR2D,
The portability in CFD code can be determined based on the performance metric by Pennycook\cite{pennycook2019implications}, the metric is expressed in a single value $P(a,p,H)$ as \cref{eq11} show, including a given application $a$, a given problem $p$, the computing platforms $H$, which is the harmonic mean of performance efficiency $e_i(a,p)$ in Harrel\cite{reguly2020productivity}: 
\begin{equation}
\label{eq11}
\Phi(a,p,H) =
\left\{\begin{matrix}
\frac{|H|}{\sum_i\in H\frac{1}{e_i(a,p)} }\ ,\forall i\in H\\
0 \ ,otherwise\\
\end{matrix}\right.
\end{equation}
\end{comment}

We implement the roofline model to give some performance insights into memory bandwidth and floating-point operations. 
Roofline\cite{williams2009roofline,harris2005mapping} is a bound and bottleneck analysis function,  it chooses the minimum value of either current peak machine performance, or peak DRAM bandwidth multiply arithmetic intensity, so that it provides insights into the primary factors affecting the performance of computer systems\cite{lazowska1984quantitative}.

\begin{comment}
	the exact formula is:
	\begin{equation}
	\begin{split}
	Flops/sec= Min(peak\ machine\ performance,\\
	peak\ DRAM\ bandwidth * arithmetic\ intensity )
	\end{split}
	\end{equation}
\end{comment}

%Performance portability metric measures the degree of achieved performance for an application on different computing platforms, but the workload is not under consideration for different computing platforms.
To measure the workload during the porting process, we apply productivity metric to quantify how efficiently HOUR2D code is developed with different programming models\cite{harrell2018effective}, specifically the "code divergence" is the average of the pairwise distances between the applications in different code versions A,
\begin{equation}
\label{eq13}
D(A)=\binom{|A|}{2}^{-1} \sum d(a_i,a_j) , ( \{ a_i,a_j \}\subset A)
\end{equation}
in which $d_{a,b}$ originates from the same code in the number of Source Line of Code(SLOC)\cite{nguyen2007sloc} normalized to the smaller application, the detailed equation is shown in \cref{eq12}.
\begin{equation}
\label{eq12}
d(a,b)= \frac{|SLOC(a)-SLOC(b)|}{min(SLOC(a),SLOC(b))}
\end{equation}

\subsection{Performance Results}

We use the complete grids to verify the correctness of output flow field, two intuitive shapes are choosen to display the stable flow field as \cref{fig7} shows.
The subfigure \cref{fig7}a displays the final denisty of NACA airfoil illustration, the various color refers to differentiated values.
While \cref{fig7}b is the pressure of flow around double cylindrical, the flow clusters clusters in front of the first circular and forms relatively high pressure, the reversed side of the circular forms low pressure.

\begin{figure}
	\centerline{\includegraphics[width=0.35\textwidth]{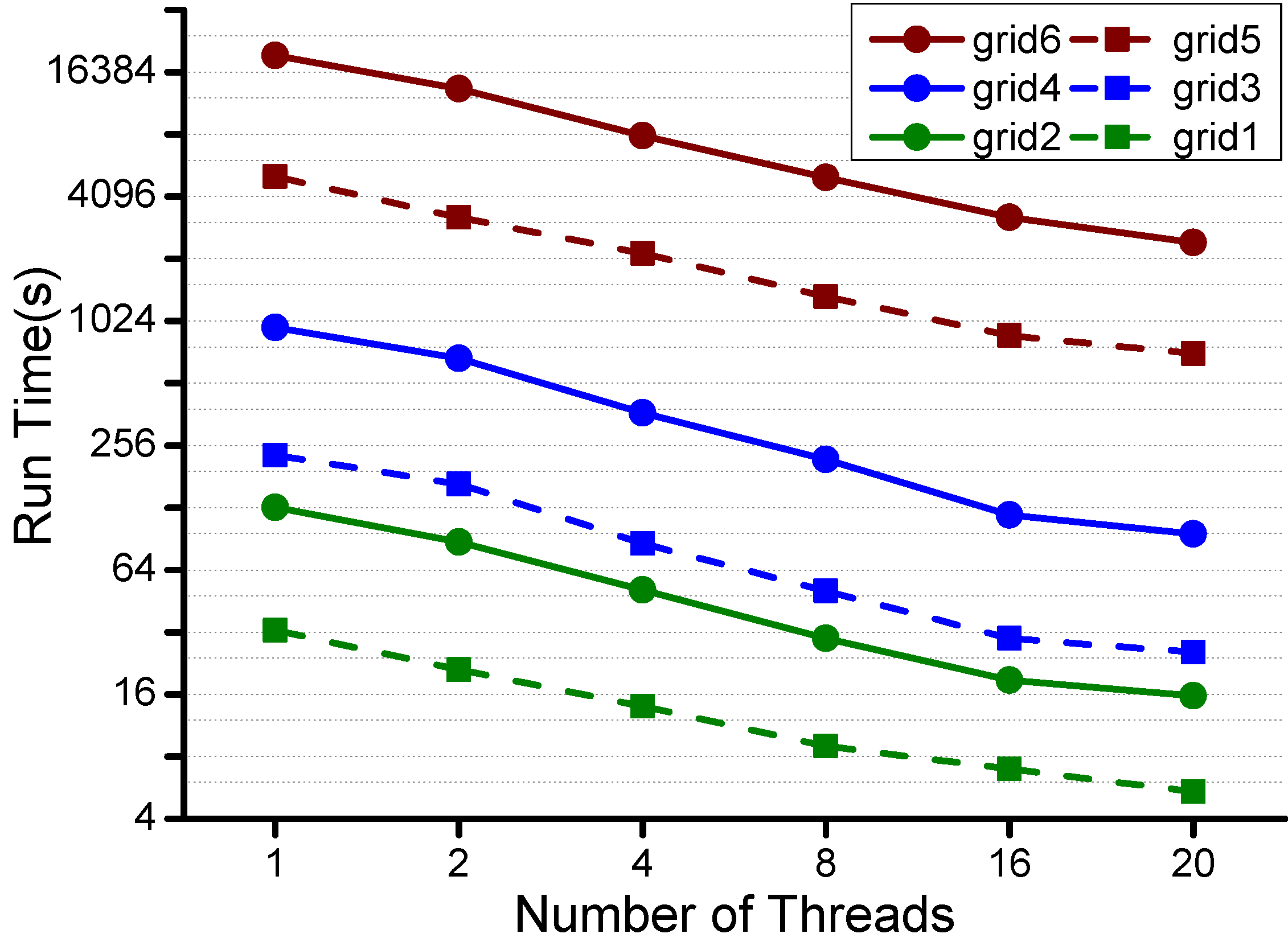}}
	\caption{ The run time of OpenMP-based code with six different grid meshes }\label{fig8}
\end{figure}

We choose OpenMP-based code as a benchamrk to assess the performance of other two languages.
Thread scalability is first evaluted with first six computational grids for the convenience of comparision, the run time is displayed in \cref{fig8}, and the number of thread varies from one to twenty.
It can be inferred from the figure that the code obtains considerable scalability as the growth rate of the running time is roughly the same as that of the grid size. 
Besides, twenty-thread parallelization achieves a speedup of 6.0x - 8.1x comparing with serial execution according to different grids, the different speedup of grid may be affected by several reasons like insuffcient parallelism for small-scale grid, memroy access exacerbation as grid size increases and so on.

\begin{figure}[htbp]
	\centerline{\includegraphics[width=0.35\textwidth]{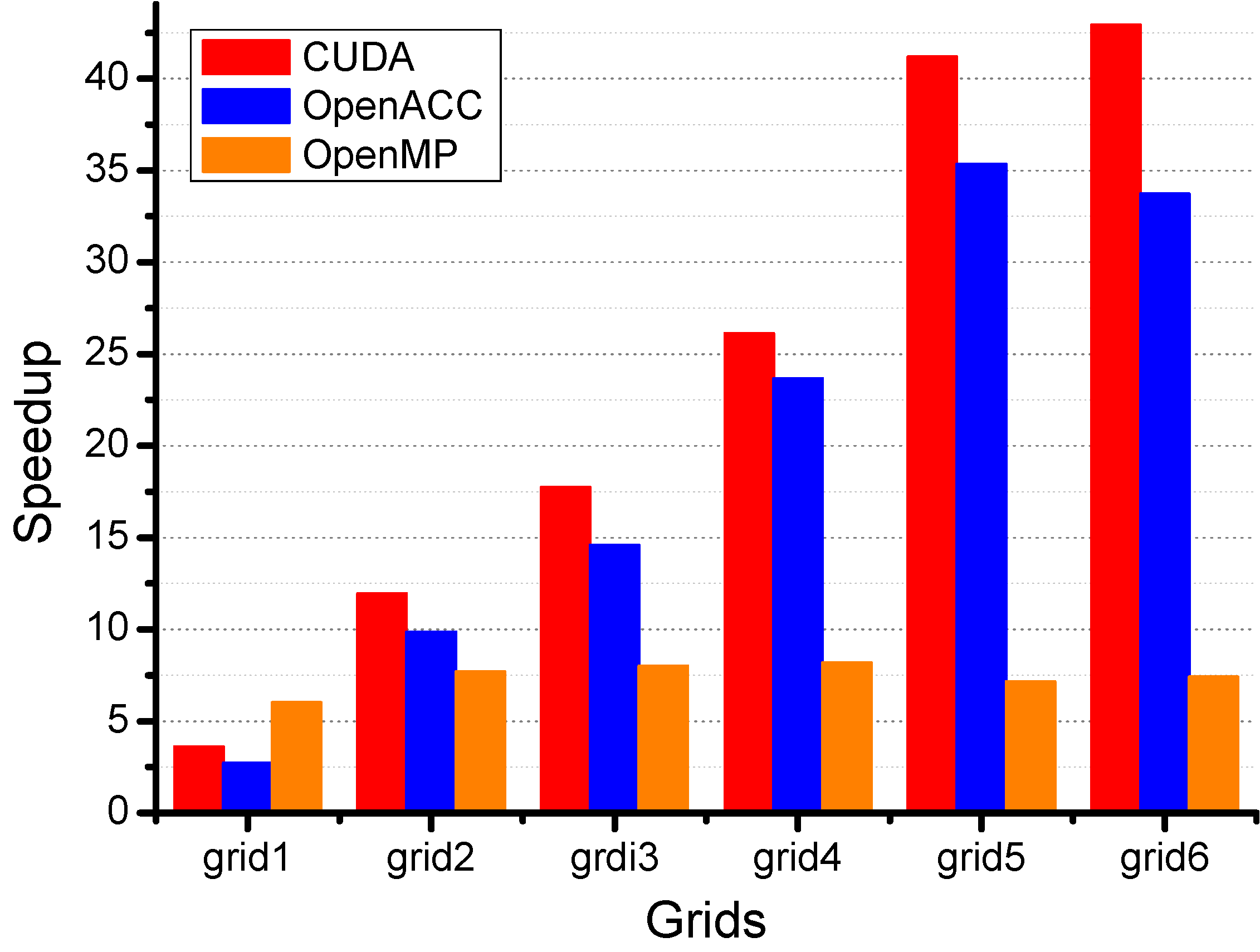}}
	\caption{ The speedup of the OpenMP, CUDA, OpenACC based code comparing with serial execution of original code. }\label{fig9}
\end{figure}

The comparisons for three languages is applied after evaluating OpenMP code base, we analyze the speedup based on the serial execution for first six grids as \cref{fig9} shows.
The colored columnars are presented under the same optimiation conditions, grid renumbering and mixed precision strategy are not included.
The GPU-based code shows relatively lower speedup than OpenMP in small-scale grids, it's due to the fact that the grid data is insufficient in parallelism gain to cover the data movement cost.
As the grid size increases, GPU-based code shows priority to OpenMP's, particularly, CUDA is much better than OpenACC owe to the fine-tuning of thread scheduling, better memory hierarchy control, and some memory access optimizations.
Generally, OpenACC's speedup reaches $75\%$ to $90\%$ of CUDA's, besides, UM approach is enabled in OpenACC-based code.
The CUDA, OpenMP, OpenACC based code obtains a maximal speedup of 42.9x, 35.3x, and 8.1x among six grids comparing with serial execution, respectively.

\begin{figure}[htbp]
	\centerline{\includegraphics[width=0.35\textwidth]{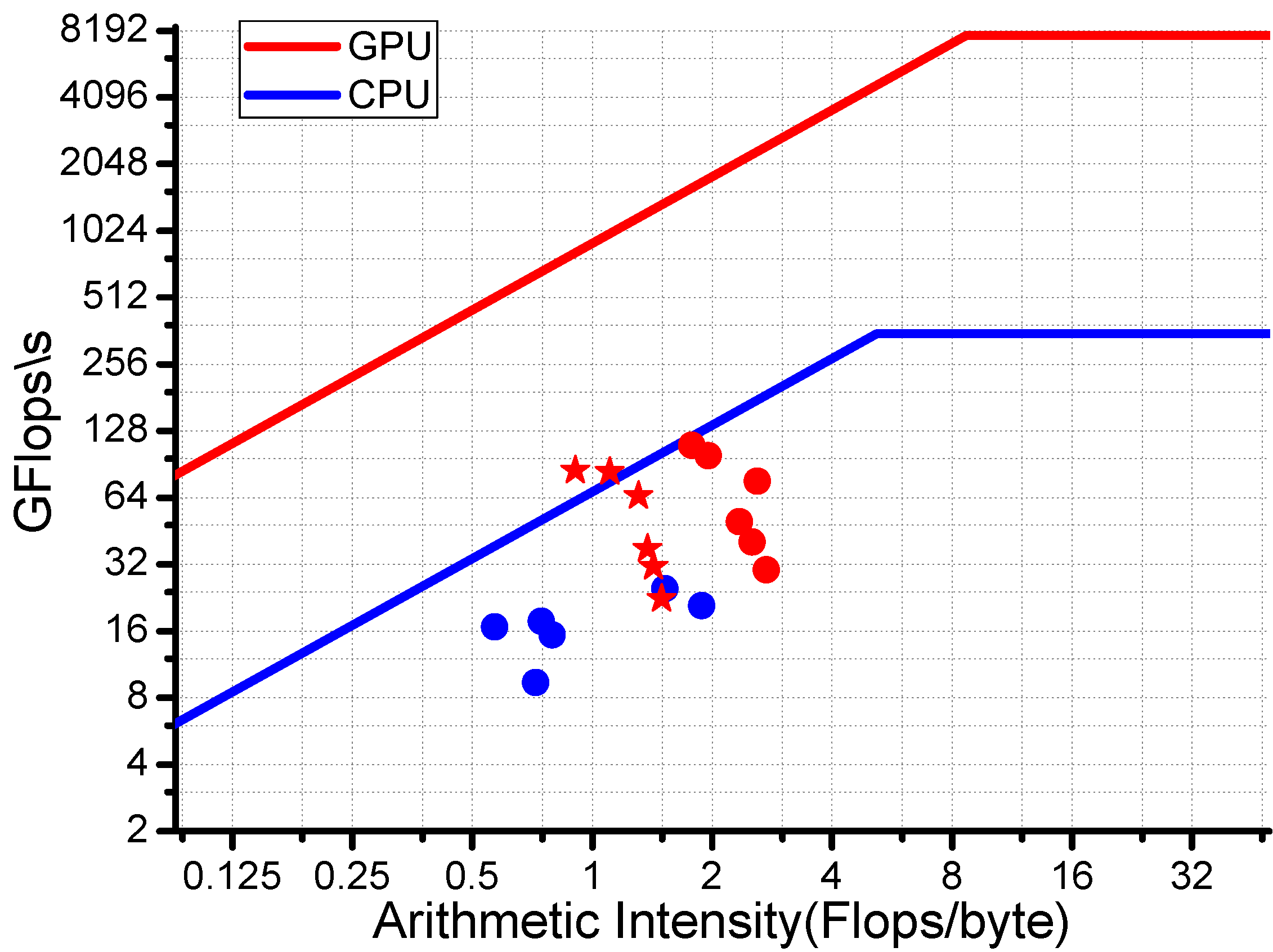}}
	\caption{ The illustration of roofline performance model for three programming models }\label{fig10}
\end{figure}

To further explore the computing performance, we build the roofline model of all three programming models to find the possible bottleneck with floating-point operations and memory access, the execution details in the CPU are collected by Intel Vtune Profiler, while the Nvidia Nsight Profiler is utilized in GPU.
The final chart is shown in \cref{fig10}, the abscissa is the arithmetic intensity which represents the amount of data required(transported from DRAM to cache ) for each floating-point operation,
and the ordinate is the double-precision floating-point operations achieved per second,
besides, the red and blue attributes belong to V100 GPU and Intel Xeon E5 CPU respectively, the solid line is the theoretical limit of the hardware unit, and the scatter points are the measured values through our experiments, moreover, the red dot shows the execution of CUDA-based code while the red star belongs to OpenACC, the different scatter points of the same color display the analysis of different grids.

We can infer from \cref{fig10} that the HOUR2D executions of all three models are faced with memory access bottleneck even if some memory optimization methods are applied, especially for CUDA-based code when grid size extends, this may be the increasing deterioration of frequent access to multi-dimensional array for element and cell integral, and the most important, non-consecutive memory access caused by the unstructured grid which is generally acknowledged as a tough issue for HPC researchers in CFD field.
What's more, the achieved Flops can be ranked as CUDA > OpenACC > OpenMP due to the abundant thread concurrency in GPU, particularly for the larger grids.

\subsection{ Productivity Evaluation }

Although the CUDA-based code shows the absolute performance advantage over OpenMP and OpenACC, meanwhile CUDA certainly consumes the most effort on code tunning.
So we use quantitative metric detailed in \cref{eq12} to further evaluate our workload as \cref{tab6} shows, in which the second column is the porting workload from $desktop$ version to current code, and the $desktop$ is taken as the benchmark version with 2900 total source lines.
The metric $d_{(current, desktop)}$ displays the relative workload porting from the desktop version to the current version as the \cref{eq11} shows, the fourth and fifth row stands for the situation of whether to use UM method or not, the bigger metric $d$ represents more porting workload and lower productivity.
So under the basis of OpenMP-based code, the workload of the CUDA is 8.2 times larger and the OpenACC using UM is 2.7 times larger, however, this quantified number has an underlying assumption that the development difficulty of different programming models keeps consistent.
\begin{table}[htbp]
	\centering
	\caption{ Portability evaluations when porting from desktop code to three programming models }
	%\vspace{0.1cm}
	\small
	\begin{tabular}{ p{3cm}p{0.9cm}p{2.8cm} }
		\toprule
		\makecell[c]{Current Version}   & \makecell[c]{ SLOC }   &   \makecell[c]{$d_{current,desktop}$(\%)}     \\
		\hline
		\makecell[c]{$OpenMP$}  & \makecell[c]{34}     & \makecell[c]{1.17}    \\
		\makecell[c]{$CUDA$}    & \makecell[c]{280}    & \makecell[c]{9.67}    \\
		\makecell[c]{$OpenACC_{UM}$} & \makecell[c]{94}     & \makecell[c]{3.24}    \\
		\makecell[c]{$OpenACC_{nonUM}$} & \makecell[c]{161}     & \makecell[c]{5.56}    \\		
		\bottomrule
	\end{tabular}
	\label{tab6}
\end{table}

%\addtolength{\rightmargin}{0.392cm}

The $relative\  effort\  time\  productivity(RDTP)$ concept connects the performance and productivity which is $\Psi_{relative}$=speedup/relative effort, it's useful under the condition of the same optimization level for three programming models, so the sort through $\Psi$ according to the metric above is: $\Psi_{OpenMP}>\Psi_{CUDA}>\Psi_{OpenACC}$. 
The OpenMP performs best because of the fairly high productivity, and CUDA is better than OpenACC.
It's worth noting that $RDTP$ takes performance and productivity equally important which is not always true in reality, striking the appropriate balance between the performance and productivity relies on the ultimate use of the code\cite{Heidrich2015Software}. 
%for example, suppose the complete HOUR2D code which is integrated in practical CFD solver, then the optimal performance regardless of productivity is unsuitable for the developers, OpenACC with UM method is a good choice when develop team  

%Our evaluation metric can give a reference for larger CFD slover which follows the similar data structure and computational formats.
%\setlength{\bottom}{4.3cm}

\section{Conclusions and Future Work}

We have presented the migration of a practical high-order DG application named HOUR2D into three programming models running on two 
architectures, performance analysis, and portability are further evaluated.

We first discuss some optimization methods in the light of data structure and computing format, the grid renumbering approach is used to relieve the discontinuous memory access, then the mixed-precision method to maximize the floating-point operations.
We also capture the performance on CPU and GPU platforms with seven grids, concretely the  CUDA, OpenMP, and OpenACC based code achieves roughly a speedup of 42.9x, 8.1x, and 35.3x based on the serial execution, then we find the memory access issue by using the roofline model which needs further research, the productivity of different programming models are measured with quantified metric.

We plan to carry out our future work in two aspects. On the one hand, we improve the memory bandwidth and the floating-point efficiency of the high-order algorithm. On the other hand, we evaluate the performance portability of our algorithm in more programming models to offer a practical assessment to the CFD developers.  

%\section*{Acknowledgment}
%This work was supported by the National Numerical Wind Tunnel of China.

\vspace{12pt}
\bibliographystyle{ieeetr}
\bibliography{ref}
\end{document}